Spectral Risk Measures: Properties and Limitations

By

Kevin Dowd, John Cotter and Ghulam Sorwar[*]

Abstract

Spectral risk measures (SRMs) are risk measures that take account of user risk-aversion, but to date there has been little guidance on the choice of utility function underlying them. This paper addresses this issue by examining alternative approaches based on exponential and power utility functions. A number of problems are identified with both types of spectral risk measure. The general lesson is that users of spectral risk measures must be careful to select utility functions that fit the features of the particular problems they are dealing with, and should be especially careful when using power SRMs.

Keywords: coherent risk measures, spectral risk measures, exponential utility, power utility

JEL Classification: G15

April 18, 2008

[*] Kevin Dowd is at the Centre for Risk and Insurance Studies, Nottingham University Business School, Jubilee Campus, Nottingham NG8 1BB, UK; email: Kevin.Dowd@nottingham.ac.uk. John Cotter is at the UCLA Anderson School of Management, 110 Westwood Plaza B504, Los Angeles, California, 90095, USA, ph. 001 310 825 2247,Email: john.cotter@anderson.ucla.edu; and at the Centre for Financial Markets, School of Business, University College Dublin, Carysfort Avenue, Blackrock, Co. Dublin, Ireland; email: john.cotter@ucd.ie. Ghulam Sorwar is at Nottingham University Business School, Jubilee Campus, Nottingham NG8 1BB, UK; email: Ghulam.Sorwar@nottingham.ac.uk The authors would like to thank Carlo Acerbi and Dirk Tasche for fruitful conversations on the subject, and they thank an anonymous referee for very helpful suggestions that have much improved the paper. Dowd's contribution was supported by an Economic and Social Research Council research fellowship on 'Risk measurement in financial institutions', and he thanks the ESRC for their financial support. Cotter's contribution to the study has been supported by a University College Dublin School of Business research grant.

## 1. Introduction

One of the most interesting and potentially most promising recent developments in the financial risk area has been the theory of spectral risk measures, recently proposed by Acerbi (2002, 2004). Spectral risk measures (SRMs) are closely related to the coherent risk measures proposed a little earlier by Artzner *et al.* (1997, 1999), and share with the coherent risk measures the highly desirable property of subadditivity. More formally, if $\rho(.)$ is a measure of risk, and if *A* and *B* are any two positions, then subadditivity means that it will always be the case that $\rho(A+B) \leq \rho(A) + \rho(B)$. Subadditivity reflects the common-sense notion that individual risks typically diversify (or, at worst, do not increase) when we put risky positions together.

One of the nice features of SRMs is that they relate the risk measure to the user's risk-aversion – in effect, the spectral risk measure is a weighted average of the quantiles of a loss distribution, the weights of which depend on the user's risk-aversion. Spectral risk measures therefore enable us to link the risk measure to the user's attitude towards risk, and we might expect that if a user is more risk averse, other things being equal, then that user should face a higher risk, as given by the value of the SRM. SRMs can be applied to many different problems. For example, Acerbi (2004) suggests that they can be used to set capital requirements or obtain optimal risk-expected return tradeoffs, Overbeck (2004) discusses how they might be used for capital allocation, and Cotter and Dowd (2006) suggest that SRMs could be used by futures clearinghouses to set margin requirements that reflect their corporate risk aversion.

However the existing literature gives very little guidance on the choice of risk aversion function or on the question of what a suitable risk aversion function might entail. For instance, Szegö (2002) describes the process of multiplying coherent risk measures by an admissible risk aversion function but does not specify what an admissible risk aversion function might be. Similarly, Acerbi (2004, p. 175) calls for the identification of additional criteria to assist the risk manager in choosing an optimal risk aversion function for a portfolio, but he himself illustrates only one particular risk-aversion function – namely, an exponential one.

This paper investigates this issue further, and examines alternative SRMs based on alternative underlying utility functions. The ones considered are 'exponential SRMs' based on an exponential utility function, which are equivalent to the ones that Acerbi studied, and 'power SRMs' based on a power utility function. To our knowledge, these latter have received no attention so far in the published literature, but they are a natural object of study as the power utility function is very widely used in other contexts.

The article is organised as follows. Section 2 sets out the essence of Acerbi's theory of spectral risk measures. Section 3 examines the properties of exponential SRMs, and section 4 does the same for power SRMs. Section 5 concludes.

**2. Spectral Risk Measures**

Consider a risk measure $M_\phi$ defined by:

$$(1) \qquad M_\phi = \int_0^1 \phi(p) q_p \, dp$$

where $q_p$ is the $p$ loss quantile and $\phi(p)$ is a user-defined weighting function defined over the full range of cumulative probabilities $p \in [0,1]$ (see also Acerbi, 2002, 2004). $M_\phi$ defines the class of quantile-based risk measures, and each individual risk measure in this class is characterized by its own particular weighting function $\phi(p)$.

Two well-known members of this class are the VaR and the Expected Shortfall (ES):
- The VaR at the $\alpha$ confidence level is:

(2) $$VaR_\alpha = q_\alpha$$

The VaR places all its weight on the $\alpha$ quantile, i.e., the VaR weighting function $\phi(p)$ is a Dirac delta function that gives the outcome $p = \alpha$ an infinite weight and gives every other outcome a weight of zero.

- The ES at the confidence level $\alpha$ is the average of the worst $1-\alpha$ losses, viz.:

(3) $$ES_\alpha = \frac{1}{1-\alpha}\int_\alpha^1 q_p\, dp$$

The ES weighing function $\phi(p)$ gives all tail quantiles the same weight of $\frac{1}{1-\alpha}$ and gives non-tail quantiles a weight of zero.

Thus, the VaR is based on a degenerate weighing function and the ES is based on a simple step weighting function. It can also be shown neither of these risk measures makes any allowance for the user being risk-averse (see, e.g., Grootveld and Hallerbach, 2004, pp. 34-35).

A user who is risk-averse might prefer to work with a risk measure that takes account of his/her risk aversion, and this takes us to the class of spectral risk measures (SRMs). In loose terms, an SRM is a quantile-based risk measure that takes the form of (1) where $\phi(p)$ reflects the user's risk aversion. More precisely, following Acerbi, we can define SRMs as the subset of $M_\phi$ that satisfy the following properties of nonnegativity, normalisation and increasingness:

*P1. Nonnegativity*: $\phi(p) \geq 0$.

*P2. Normalisation*: $\int_0^1 \phi(p)dp = 1$.

*P3. Increasingness*: $\phi'(p) \geq 0$.[1]

The first coherent condition requires that the weights are nonnegative and the second requires that the probability-weighted weights should sum to 1, but the key condition is the third one. This condition requires that the weights attached to higher losses should be no less than the weights attached to lower losses, and is intended to reflect user risk-aversion.

---

[1] See Acerbi (2002, 2004). Strictly speaking, Acerbi's P3 is a decreasingness condition, but he is dealing with distributions in which loss outcomes are given negative rather than positive values. However, this difference is insubstantial and our conditions P1-P3 are equivalent to his.

However, a drawback with property P3 is that it does not rule out risk-neutral risk measures from the set of SRMs. For instance, the ES would qualify as an SRM under P3, and we have already seen that the ES does not accommodate user risk aversion. To rule out such cases, we replace P3 with the following slightly stronger condition:

*P3'. Strict increasingness*: $\varphi'(p) > 0$.

Condition P3' ensures that the weight $\phi(p)$ rises with *p*. In 'well-behaved' cases, we would expect the weights to rise smoothly, and to rise more rapidly for users who are more risk-averse.

A risk measure that satisfies these properties is attractive not only because it takes account of user risk-aversion, but also because such a risk measure is known to be coherent (see Acerbi, 2004, Proposition 3.4). Thus, SRMs have the various attractions of coherent risk measures (and especially subadditivity).

There still remains the question of how to specify $\phi(p)$, and perhaps the most natural way to obtain $\phi(p)$ is from the user's utility function (see also Bertsimas *et al.*, 2004).

## 3. Exponential Spectral Risk Measures

This requires us to choose a utility function, and a natural choice is the following exponential utility function defined over outcomes *x*:

$$U(x) = -e^{-kx} \tag{4}$$

where $k > 0$ is the Arrow-Pratt coefficient of absolute risk aversion (ARA). The coefficients of absolute and relative risk aversion are:

$$R_A(x) = -\frac{U''(x)}{U'(x)} = k \tag{5a}$$

(5b) $$R_R(x) = -\frac{xU''(x)}{U'(x)} = xk$$

To obtain our weighting function, we set

(6) $$\varphi(p) = \lambda e^{-k(1-p)}$$

where $\lambda$ is an unknown positive constant.[2] This clearly satisfies properties 1 and 3, and we can easily show (by integrating $\phi(p)$ from 0 to 1, setting the integral to 1 and solving for $\lambda$) that it satisfies 2 if we set

(7) $$\lambda = \frac{k}{1-e^{-k}}$$

Hence, substituting (7) into (6) gives us the exponential weighting function corresponding to (4):

(8) $$\phi(p) = \frac{ke^{-k(1-p)}}{1-e^{-k}}$$

This weighting function is illustrated in Figure 1 for two alternative values of the ARA coefficient, $k$. Observe that this weighting function has a nice shape and rises exponentially with $p$. In addition, for the higher $p$ values associated with higher losses, the weights are higher and the rate of increase of $\phi(p)$ is higher, the greater the value of the ARA coefficient.

**Insert Figure 1 here**

---

[2] A weighting function of the form given in (6) is a natural choice for an exponential utility function, as it reflects the structure of the utility function. We do not assert that this weighting function is unique, but we have not been able to find any alternative that also fits the necessary criteria.

The SRM based on this weighting function, the exponential SRM, is then found by substituting (8) into (1), viz.:

$$(9) \quad M_\phi = \int_0^1 \phi(p) q_p \, dp = \frac{k}{1-e^{-k}} \int_0^1 e^{-k(1-p)} q_p \, dp$$

The value of the risk measure can then be found using numerical integration.

The first question of interest is how the SRM changes with the coefficient of risk aversion. As proven in the Appendix, it is not possible to say that $\partial M_\phi / \partial k > 0$ for all possible distributions, but some plots of the SRM against $k$ for various illustrative distributions are shown in Figure 2. The distributions illustrated are standard normal, Cauchy, standard uniform, a beta with a right-hand skew and a Gumbel, a form of extreme-value distribution. In every case, the SRM rises with $k$ in a 'well-behaved' manner, and the fact that such different distributions produce qualitatively similar plots suggests that $\partial M_\phi / \partial k > 0$ must commonly though not universally hold. Some illustrative values of the exponential SRM under these alternative loss distributions are given in Table 1. So, for example, if we set $k = 5$, the spectral risk measure under standard normality is 1.080, but if we increase $k$ to 25, the same measure rises to 1.945.

**Insert Table 1 here**
**Insert Figure 2 here**

However, the exponential SRM also has the rather odd property that the value of the risk measure approaches the mean of the loss distribution in the limit as the value of $k$ goes to zero, viz.:

$$(10) \quad M_\phi \to \int_0^1 q_p \, dp \text{ as } k \to 0$$

This property is also proved in the Appendix. This is a rather strange property, and one that also goes against the fairly natural expectation that a 'sensible' risk measure should always be sensitive to conditions such as market volatility. Note, too, that this property holds for any loss distribution.

Finally, there is the question of whether the exponential utility function provides a good description of empirically plausible risk aversion. The answer here is mixed:

- On the one hand, the exponential utility function implies that the coefficient of absolute risk aversion is constant and the coefficient of relative risk aversion increases with wealth (see (5) above). However, the generally accepted stylised facts are that real-world agents exhibit decreasing absolute risk aversion (because a rich person would usually require a smaller premium to accept a given gamble than a poorer one) and constant relative risk aversion (because society now is much wealthier than it used to be, but there seems no obvious connection between Gross Domestic Product and observable risk premiums). Thus, the absolute and relative risk aversion properties of the exponential do not match what we think we observe in the real-world, and this suggests that the exponential might not always be appropriate.
- On the other hand, the theoretical work of Buhlmann (1980) shows that, under weak conditions, all equilibrium prices are locally like the ones that would arise if agents had exponential utilities but where risk aversion is also dependent on net wealth (see also Wang (2003)). This suggests that the exponential utility function might be plausible in circumstances where we were dealing with a hypothetical 'representative agent' and were trying to infer this agent's risk-aversion parameters from financial market prices.

Users of exponential SRMs therefore need to make sure that they use them in circumstances that are empirically plausible.

**4. Power Spectral Risk Measures: $\gamma < 1$**

We can also obtain SRMs based on other utility functions, and a popular alternative to the exponential utility function is the power utility function:

(11) $$U(x) = \frac{x^{1-\gamma} - 1}{1-\gamma}$$

for some positive parameter $\gamma > 0$, and where

(12) $$U(x) = \ln(x)$$

in the limiting case where $\gamma = 1$. Its coefficients of absolute and relative risk aversion are:

(13a) $$R_A(x) = -\frac{U''(x)}{U'(x)} = \frac{\gamma}{x}$$

(13b) $$R_R(x) = -\frac{xU''(x)}{U'(x)} = \gamma$$

Thus, the power utility function has a constant coefficient of relative risk aversion equal to our parameter $\gamma$. This function therefore belongs to the family of Constant Relative Risk Aversion (CRRA) utility functions.

Our next task is to specify the weighting function, and one choice is the following:

(14) $$\varphi(p) = \lambda \frac{(1-p)^{\gamma-1}}{1-\gamma}$$

where $\lambda$ is another unknown constant.[3] We can easily show that this function satisfies property 2 if we set:

(15) $$\lambda = \gamma(1-\gamma)$$

---

[3] Apropos note 2, a weighting function of the form given in (14) is a natural choice for the power utility function with $\gamma < 1$ - and the same goes for (20) or (21) below for the power utility function with $\gamma > 1$ - as it reflects the structure of the utility function. And, as with the earlier exponential case, we do not assert that this weighting function is unique, but are unable to find any alternatives that also satisfy the necessary criteria.

Substituting (15) into (14) then gives:

(16) $$\varphi(p) = \gamma(1-p)^{\gamma-1}$$

It is then obvious that property 1 always holds, and property 3' holds if $\gamma < 1$. We note at this point that this latter restriction might be a problem, because there is no *a priori* reason why $\gamma$ should be less than 1, and there may be circumstances where we are dealing with $\gamma$ values that exceed 1 (see, e.g., Dowd *et al.*, 2008). We shall come back to this issue presently.

To investigate its properties, the power weighting function (16) is plotted in Figure 3 for illustrative $\gamma$ values equal to 0.7 and 0.9. This shows that, as we move right, the higher RRA-$\phi(p)$ curve is initially higher than the lower RRA-$\phi(p)$ curve, but then falls below it once $p$ reaches a certain level. This tells us that with *higher* risk aversion, relatively *more* weight is placed on the *lower* losses and relatively less weight is placed on the *higher* losses! This is clearly odd, even though the $\phi(p)$ function satisfies properties 1-3' set out above.

**Insert Figure 3 here**

The resulting risk measure (obtained by substituting (16) into (1)) is then

(17) $$M_\varphi = \int_0^1 \varphi(p) q_p \, dp = \int_0^1 \gamma(1-p)^{\gamma-1} q_p \, dp$$

and again the values of the risk measure can be found using numerical integration.

This SRM satisfies the following two properties which are sufficiently obvious that they do not need any explicit proof:

(18) $$M_\phi \to 0 \text{ as } \gamma \to 0$$

(19) $$M_\phi \to \int_0^1 q_p \, dp \text{ as } \gamma \to 1$$

The first property, (18), indicates that the PSRM approaches a 'singular point' of 0 as $\gamma \to 0$. This implies that the PSRM is totally insensitive to market conditions and to the form of the loss distribution function in the limit when $\gamma = 0$. The second property, (19), tells us that the value of the PSRM approaches the mean of the loss distribution as $\gamma \to 1$, i.e., we have a 'near singular point' at $\gamma = 1$. This implies that the PSRM becomes completely insensitive to the market volatility or to the form of the loss distribution in the limit as $\gamma \to 1$. Thus, as we move from $\gamma = 0$ towards $\gamma = 1$, the PSRM always starts at one value, 0, and always ends at another value, the mean of the loss distribution, and this is the case for all possible loss distributions. From the risk measurement point of view, these 'singular' and 'near-singular' points are bizarre features that cast further doubt on the suitabilty of PSRMs as risk measures.

To illustrate their properties further, Figure 4 shows plots of the power SRMs (PSRMs) against $\gamma$ and Table 2 gives some numerical values, each obtained under the same alternative illustrative loss distributions as before (i.e., that losses are respectively standard normal, Cauchy, standard uniform, beta and Gumbel distributed). In each case, the SRM starts at zero (as it must), then quickly rises, peaks and falls back down. Thus, once it passes its peak, the SRM subsequently *falls* as the user becomes *more* risk-averse. A risk measure that falls as the user becomes more risk-averse is, to say the least, rather odd.

**Insert Figure 4 here**
**Insert Table 2 here**

Thus, we have a spectral risk measure that satisfies Acerbi's conditions, and yet the weighting function and resulting risk measure are manifestly 'badly-behaved'. Properties 1 to 3 (or 1 to 3') are clearly not sufficient to ensure that we get a 'well-behaved' risk aversion function or a 'well-behaved' SRM, at least not with power utility and $\gamma < 1$.

**5. Power Spectral Risk Measures: $\gamma > 1$**

We turn now to seek a weighting function for a power utility function compatible with $\gamma > 1$. Following Dowd *et al* (2008), we now postulate an alternative weighting function that also has power utility properties, viz.:

$$(20) \qquad \varphi(p) = \lambda p^{\gamma - 1}$$

where $\lambda$ is again an unknown constant. It is easily demonstrated that (20) satisfies property 2 if we set $\lambda = \gamma$. Our weighting function then becomes:

$$(21) \qquad \varphi(p) = \gamma p^{\gamma - 1}$$

and it is easily shown that this function always satisfies property 1 and satisfies property 2 provided $\gamma > 1$. Accordingly, we now impose this restriction and assume $\gamma > 1$.

The power weighting function (21) is plotted in Figure 5 for illustrative $\gamma$ values equal to 1.5 and 5. In each case, the weighting function starts at 0 for $p = 0$ and ends up equal to the relevant value of $\gamma$. The two cases differ, however, in that $\phi(p)$ rises at a decreasing rate with $p$ if $\gamma < 2$; but if $\gamma > 2$, then $\phi(p)$ rises at a increasing rate. Nonetheless, the shapes of both curves are still 'well-behaved'.

**Insert Figure 5 here**

The resulting PSRM is obtained by substituting (21) into (1), viz.:

$$(22) \qquad M_\phi = \int_0^1 \phi(p) q_p dp = \int_0^1 \gamma p^{\gamma-1} q_p dp$$

As with the $\gamma < 1$ PSRM, the sign of $\partial M_\phi / \partial \gamma$ for the $\gamma > 1$ PSRM is theoretically ambiguous. (This claim is also proven in the Appendix.) To illustrate their properties, Figure 6 shows plots of these PSRMs against $\gamma$, and Table 3 gives some numerical examples, each based on our earlier set of alternative loss distributions. In each case considered, the PSRM rises with $\gamma$ but at a decreasing rate: in this respect (and for at least these particular loss distributions), the $\gamma > 1$ PSRMs seem to behave more like the exponential SRMs rather than their $\gamma < 1$ relatives.

**Insert Figure 6**
**Insert Table 3**

In addition, it is immediately apparent that the $\gamma > 1$ PSRM always goes to the mean loss as $\gamma$ declines to 1, viz:

$$(23) \qquad M_\phi \to \int_0^1 q_p \, dp \text{ as } \gamma \to 1$$

Thus, the PSRM for $\gamma > 1$ has a 'near singular point' at $\gamma = 1$ where it is totally insensitive to market volatility or to the form of the loss distribution.

If we compare these results with the earlier power results for the $\gamma < 1$ case, we can see that these are better – because the shapes of the weighting function curves are much better, because the SRM rises with $\gamma$ (at least with the illustrative distributions we considered) and because we have only one singular point instead of a singular point and a near-singular point – but this singular point is still a problem.

There are also other problems when we consider the full possible range of values that $\gamma$ might take, i.e., when we consider the full range $\gamma > 0$. One problem is that we have to apply two different kinds of power SRM depending on whether $\gamma$ is less than 1 or greater than 1. This is clearly unsatisfactory, but we are unable to find any 'generic' PSRM that can be applied to both $\gamma < 1$ and $\gamma > 1$.

Now consider a hypothetical agent whose risk aversion changes over time. More precisely, let us suppose that this agent starts off with a $\gamma$ that is initially 0, but then rises over time, breaches the $\gamma = 1$ boundary and then continues to rise. Putting our results together, we then end up with the following story: our agent starts with a PSRM of 0 and the PSRM will approach the mean of the loss distribution as $\gamma \to 1$. (With the specific distributions we considered, this involved the PRSM rising, then peaking and falling back as $\gamma \to 1$, but other behaviour may be possible for other distributions, although in every case the PSRM must start at 0 and approach the mean loss as $\gamma \to 1$.) It then passes through the $\gamma = 1$ 'black hole' point and rises thereafter.

If this sounds strange, now consider the same history viewed from the perspective, not of the value of the PSRM, but of the PSRM's sensitivity to market conditions. The story now goes as follows: at first the PSRM has absolutely no sensitivity to either the market mean or volatility; it then gradually becomes sensitised to these factors, but as $\gamma$ gets larger and starts to approach 1 it loses its sensitivity to the market volatility; it then passes through the 'black hole' at $\gamma = 1$; however, as $\gamma$ continues to rise, its sensitivity to market volatility starts to grow again.

We would suggest that such bizarre properties seriously undermine the suitability of SRMs based on power utility functions.

**5. Conclusions**

This paper has examined spectral risk measures based on exponential and power utility functions. We find that the exponential utility function leads to risk-aversion functions and spectral risk measures with some intuitive properties. They are admittedly subject to the drawback that the value of the exponential SRM always goes to the mean loss as the coefficient of absolute risk aversion goes to zero, but even with this restrictive property, one could imagine users choosing to adopt the exponential SRM because of its 'nice' features, and an example would be a futures clearinghouse that might choose an SRM to determine margin requirements (Cotter and Dowd, 2006). The selection of the exponential utility function and the value of the ARA parameter would then be matters of clearinghouse corporate policy.

When dealing with power utility functions, on the other hand, we find two quite different cases depending on whether the coefficient of relative risk aversion, $\gamma$, is less than 1 or greater than 1. In the former case, the weighting functions $\phi(p)$ have counter-intuitive properties, and a plot of the SRM against $\gamma$ will show the risk measure starting from 0 before approaching the mean loss as $\gamma \rightarrow 1$. For its part, the $\gamma > 1$ always starts from the mean loss at the point where $\gamma = 1$. In neither case can we rule out the possibility that the risk measure falls as the coefficient of risk aversion rises, but in the illustrative distributions we examined, we found cases where this occurred only where $\gamma < 1$. In addition, the fact that we have two different types of power SRM corresponding to two mutually exclusive ranges of $\gamma$ is another limitation of power SRMs.

In short, our investigation reveals that SRMs can have some curious and surprising properties – some of which undermine their usefulness for practical risk management – and this is especially the case for power SRMs. The general lesson is that users of spectral risk measures must be careful to ensure that they pick utility functions that fit the features of the particular problems they are dealing with, and they should be especially careful when using power SRMs.

Finally, we reiterate two important caveats. First, the results reported in this paper were obtained using a small set of alternative loss distributions, so we cannot rule out the possibility that we might get qualitatively different results with other distributions that we have not examined. And, second, we cannot rule out the possibility that there exist alternative weighting functions compatible with the utility functions considered here – although we have no reason to suspect that such weighting functions actually exist – and that these might produce substantially different results from those reported here. Nonetheless, our results are quite revealing and give us some sense of the properties of these risk measures.[4]

---

[4] There is also another problem with all the SRMs considered here. If we examine the partial derivative of any SRM with respect to its coefficient of risk aversion, we find that these are collections of integrals all ending in $q_p dp$ terms. (Two of these are illustrated in the Appendix, and the other one is straightforward.) We can now add or subtract any fixed amount to all the

---

quantiles and if the amount chosen is large enough, the sign of the partial derivative $\partial M_\phi / \partial k$ will change. This establishes that these partial derivatives are not translationally invariant, even though the risk measures themselves are. (The risk measures are translationally invariant because they are coherent, and translational invariance of the risk measure is one of the properties of coherence: see Artzner *et al*., 1999) This is a profound problem that warrants further investigation and gives us additional grounds for concern about the properties of SRMs.

**Appendix: Proofs**

*Proof that the sign of $\partial M_\phi / \partial k$ is ambiguous.*

Differentiating (9), we obtain

$$\frac{\partial M_\phi}{\partial k} = \frac{\partial}{\partial k}\left[\frac{k}{1-e^{-k}}\right]\int_0^1 e^{-k(1-p)} q_p dp + \left[\frac{k}{1-e^{-k}}\right]\int_0^1 \frac{\partial}{\partial k}[e^{-k(1-p)}] q_p dp$$

$$= \left[1-(1-k)e^{-k}\right]\int_0^1 e^{-k(1-p)} q_p dp - \left[\frac{k}{1-e^{-k}}\right]\int_0^1 e^{-k(1-p)}(1-p) q_p dp$$

$$= (1-e^{-k})\int_0^1 e^{-k(1-p)} q_p dp + ke^{-k}\int_0^1 e^{-k(1-p)} q_p dp - \frac{k}{1-e^{-k}}\int_0^1 e^{-k(1-p)}(1-p) q_p dp$$

$$= \left(1-e^{-k}+ke^{-k}-\frac{k}{1-e^{-k}}\right)\int_0^1 e^{-k(1-p)}\left(1+\frac{k}{1-e^{-k}} p\right) q_p dp$$

Whatever the sign of this expression, we can now add or subtract any fixed amount to each of the quantiles $q_p$, and if the amount added or subtracted is large enough, this will change the sign of the expression. Hence, the sign of $\partial M_\phi / \partial k$ is ambiguous.

*Proof of (10): $M_\phi \to \int_0^1 q_p dp$ as $k \to 0$*

As $k \to 0$ in (9), $M_\phi \to \lim_{k\to\infty}\left[\frac{k}{1-e^{-k}}\right]\int_0^1 q_p dp = \int_0^1 q_p dp$

applying L'Hôspital's rule.

*Proof that the sign of $\partial M_\phi / \partial \gamma$ for $\gamma > 1$ is ambiguous*

Differentiating (22), we obtain

$$\frac{\partial M_\phi}{\partial \gamma} = \int_0^1 p^{\gamma-1} q_p dp + \int_0^1 \gamma(1-\gamma) p^{\gamma-2} q_p dp = \int_0^1 \left(p^{\gamma-1}+\gamma(1-\gamma)p^{\gamma-2}\right) q_p dp$$

As with the first proof, we can now add or subtract any fixed amount to each of the $q_p$, and if the amount added or subtracted is large enough, the sign of $\partial M_\phi / \partial \gamma$ will change. Hence, the sign of $\partial M_\phi / \partial \gamma$ must be ambiguous.

# FIGURES

## Figure 1: Exponential Weighting Functions

Notes: The Figure shows the value of the exponential weighting function (8) $\phi(p) = ke^{-k(1-p)}/(1-e^{-k})$ for values of the coefficient of absolute risk aversion, $k$, equal to 5 and 25, plotted against the cumulative probability $p$.

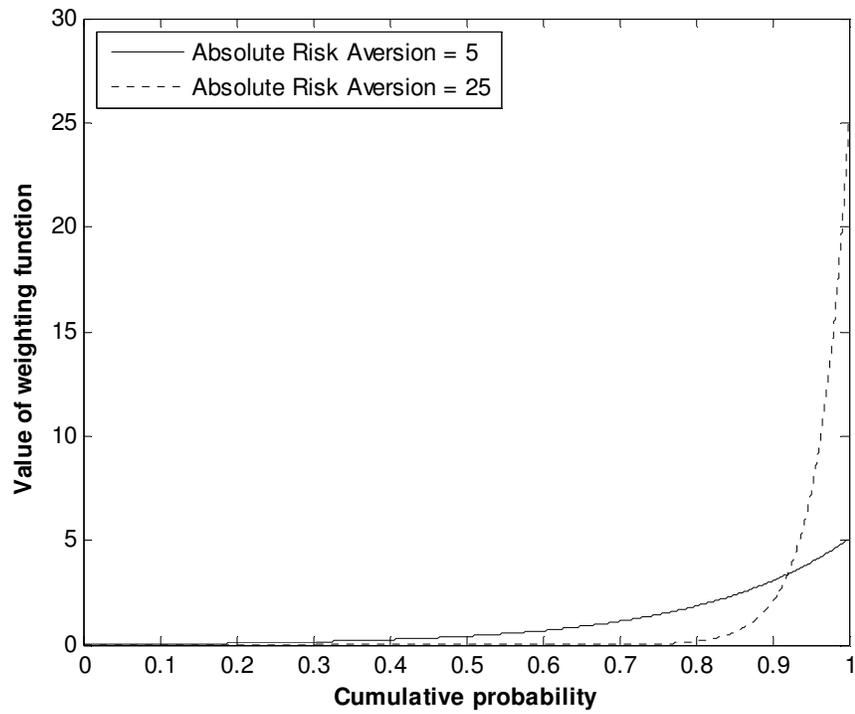

# Figure 2: Plots of Exponential Spectral Risk Measure Against the Coefficient of Absolute Risk Aversion for Various Illustrative Loss Distributions

Notes: The Figure shows the value of the exponential spectral risk measure (9) $M_\phi = \frac{k}{1-e^{-k}} \int_0^1 e^{-k(1-p)} q_p \, dp$ plotted against the coefficient of absolute risk aversion, $k$, under the alternative assumptions that losses are distributed as: standard normal, Cauchy, standard uniform, beta(2,4) and standard Gumbel. $p$ is the cumulative probability, and results are based on numerical quadrature using Simpon's rule with $p$ divided into 10,001 'slices'. The calculations were carried out using the CompEcon functions in MATLAB given in Miranda and Fackler (2002).

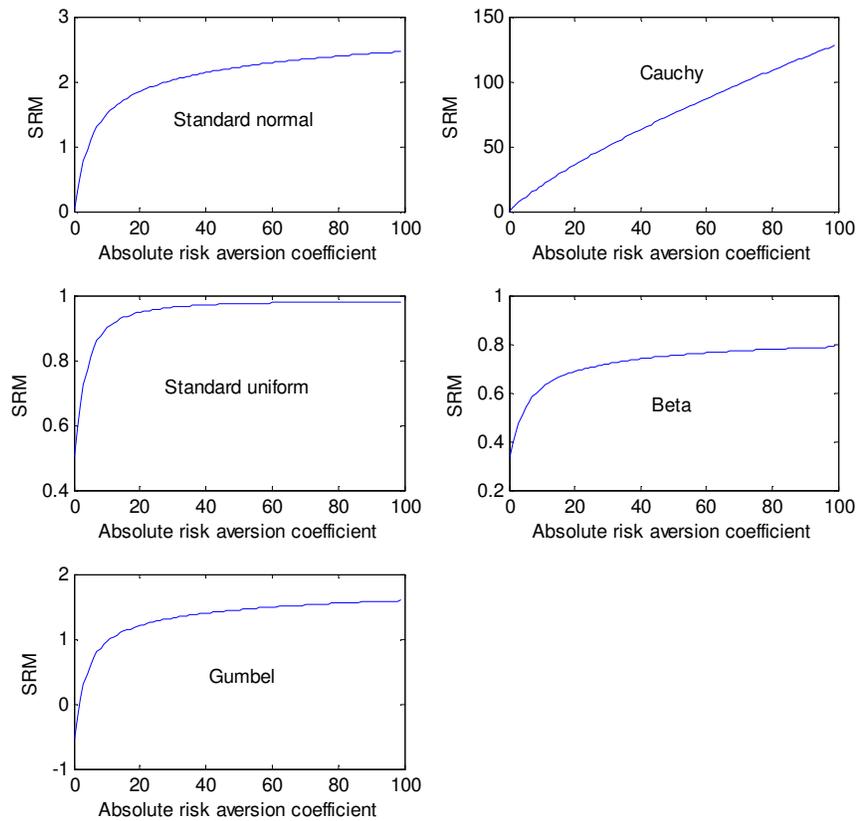

**Figure 3: Power Risk Aversion Functions:** $\gamma < 1$

Notes: The Figure shows the value of the power weighting function (16) $\varphi(p) = \gamma(1-p)^{\gamma-1}$ for the case where $\gamma$, the coefficient of relative risk aversion, is less than 1, for values of $\gamma$ equal to 0.7 and 0.9, plotted against the cumulative probability $p$.

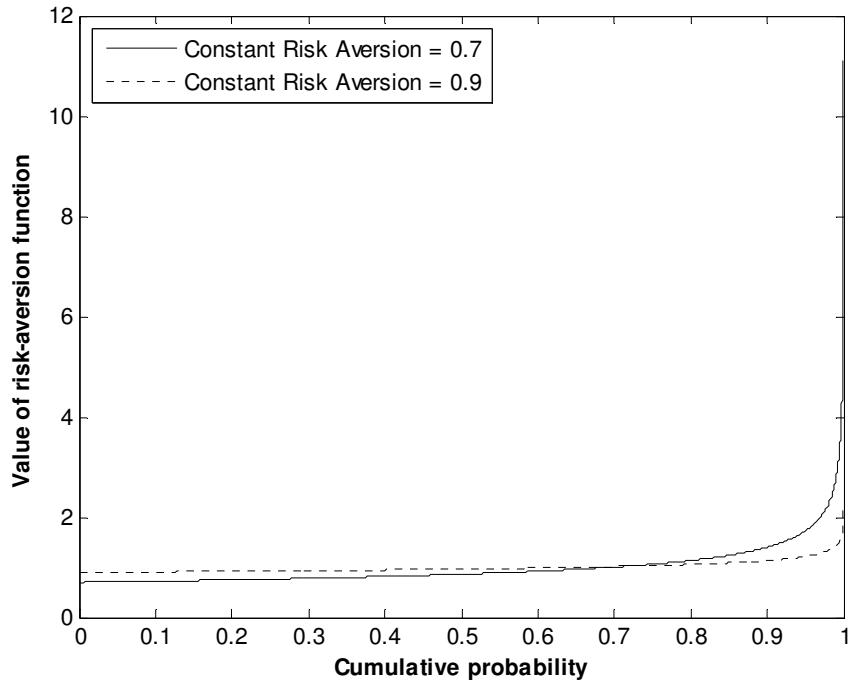

# Figure 4: Plot of Power Spectral Risk Measure Against Relative Risk Aversion: Standard Normal Loss Distribution, $\gamma < 1$

Notes: The Figure shows the value of the power spectral risk measure (17) $M_\varphi = \int_0^1 \gamma(1-p)^{c-1} q_p \, dp$ plotted against the coefficient of relative risk aversion, $\gamma$, for the case where $\gamma < 1$, under the alternative assumptions that losses are distributed as: standard normal, Cauchy, standard uniform, beta(2,4) and standard Gumbel. $p$ is the cumulative probability, and results are based on numerical quadrature using the trapezoidal rule with $p$ divided into $10,000$ 'slices'. The calculations were carried out using the CompEcon functions in MATLAB given in Miranda and Fackler (2002).

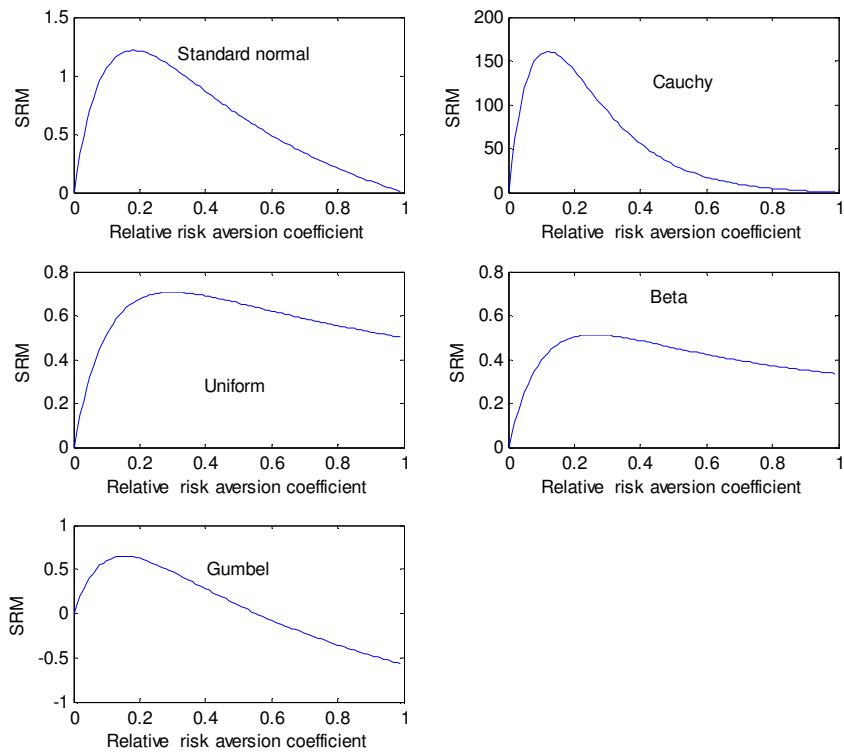

**Figure 5: Power Risk Aversion Functions:** $\gamma > 1$

Notes: The Figure shows the value of the power weighting function (21) $\varphi(p) = \gamma p^{\gamma-1}$ for the case where $\gamma$, the coefficient of relative risk aversion, exceeds 1, for values of $\gamma$ equal to 1.5 and 5, plotted against the cumulative probability $p$.

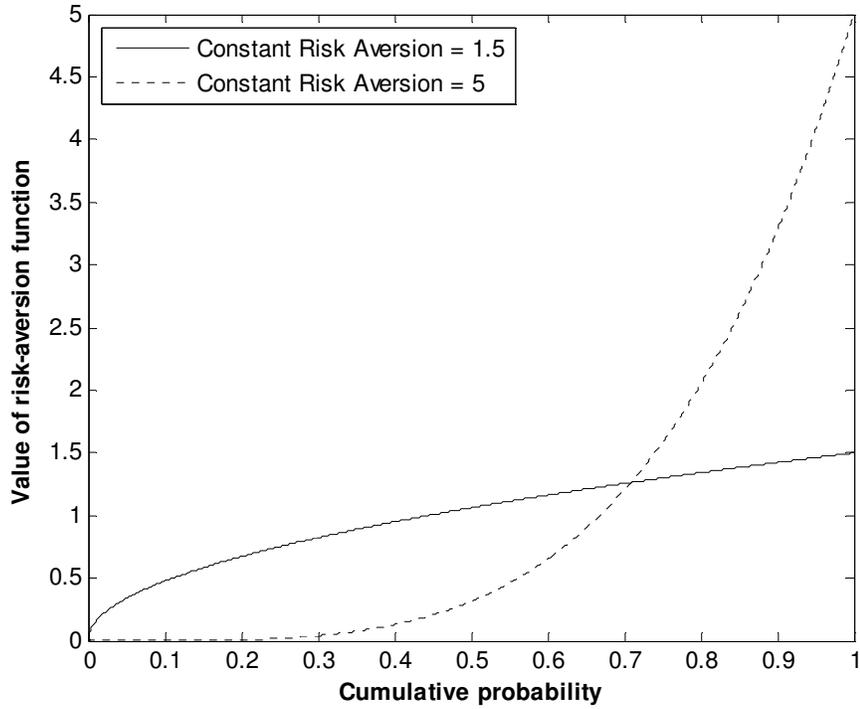

# Figure 6: Plot of Power Spectral Risk Measure Against Relative Risk Aversion: Standard Normal Loss Distribution, $\gamma > 1$

Notes: The Figure shows the value of the power spectral risk measure (22) $M_\phi = \int_0^1 \gamma p^{\gamma-1} q_p dp$ plotted against the coefficient of relative risk aversion, $\gamma$, for the case where $\gamma > 1$, under the alternative assumptions that losses are distributed as: standard normal, Cauchy, standard uniform, beta(2,4) and standard Gumbel. $p$ is the cumulative probability, and results are based on numerical quadrature using the trapezoidal rule with $p$ divided into 10,000 'slices'. The calculations were carried out using the CompEcon functions in MATLAB given in Miranda and Fackler (2002).

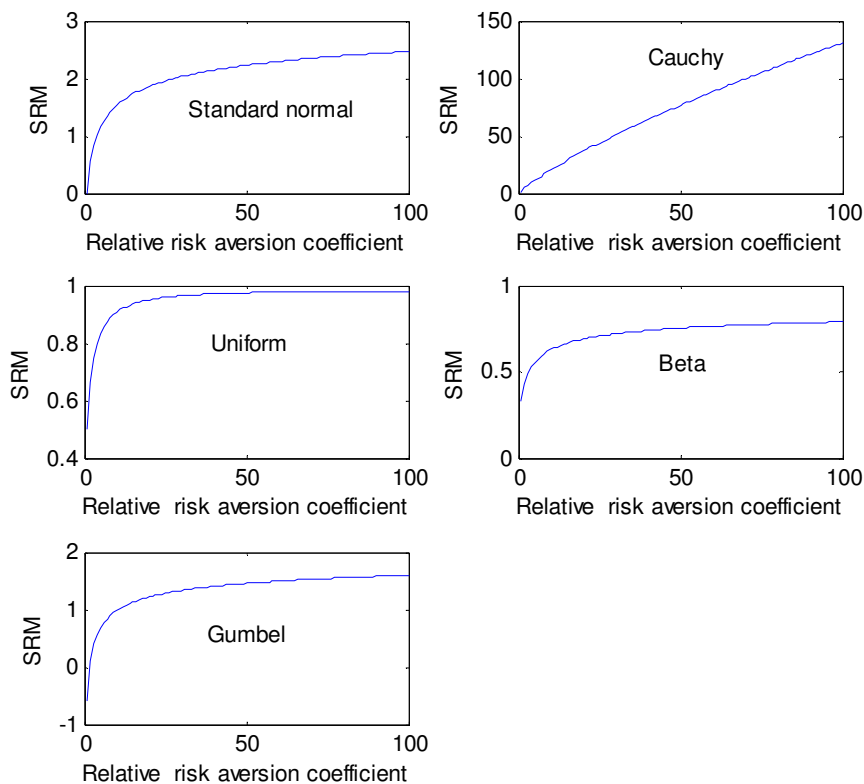

# Table 1: Values of Exponential Spectral Risk Measure under Alternative Illustrative Loss Distributions

Notes: Estimates are of exponential spectral risk measure (9) $M_\phi = \dfrac{k}{1-e^{-k}} \int_0^1 e^{-k(1-p)} q_p \, dp$ where $k$ is the coefficient of absolute risk aversion and $p$ is the cumulative probability, under the alternative assumptions that losses are distributed as: standard normal, Cauchy, standard uniform, beta(2,4) and standard Gumbel. Results are based on numerical quadrature using Simpon's rule with $p$ divided into 10,001 'slices'. The calculations were carried out using the CompEcon functions in MATLAB given in Miranda and Fackler (2002).

| Coefficient of Absolute Risk Aversion | Standard normal | Cauchy | Standard uniform | Beta | Gumbel |
|---|---|---|---|---|---|
| 1 | 0.278 | 2.341 | 0.582 | 0.384 | -0.249 |
| 5 | 1.080 | 10.955 | 0.806 | 0.538 | 0.599 |
| 25 | 1.945 | 43.166 | 0.958 | 0.706 | 1.275 |
| 100 | 2.467 | 128.930 | 0.980 | 0.789 | 1.594 |

# Table 2: Values of Power Spectral Risk Measure under Alternative Illustrative Loss Distributions: $\gamma < 1$

Notes: Estimates are of power spectral risk measure (17) $M_\varphi = \int_0^1 \gamma(1-p)^{c-1} q_p \, dp$ where $\gamma < 1$ is the coefficient of relative risk aversion and $p$ is the cumulative probability under the alternative assumptions that losses are distributed as: standard normal, Cauchy, standard uniform, beta(2,4) and standard Gumbel. Results are based on numerical quadrature using trapezoidal rule with $p$ divided into 10,000 'slices'. The calculations were carried out using the CompEcon functions in MATLAB given in Miranda and Fackler (2002).

| Coefficient of Relative Risk Aversion | Standard normal | Cauchy | Standard uniform | Beta | Gumbel |
|---|---|---|---|---|---|
| $\to 0$ | 0 | 0 | 0 | 0 | 0 |
| 0.1 | 1.062 | 157.980 | 0.514 | 0.394 | 0.597 |
| 0.5 | 0.664 | 31.707 | 0.657 | 0.454 | 0.093 |
| 0.9 | 0.096 | 1.697 | 0.526 | 0.351 | -0.472 |
| $\to 1$ | 0 | 0 | 0.500 | 0.333 | -0.576 |

# Table 3: Values of Power Spectral Risk Measure under Alternative Illustrative Loss Distributions: $\gamma > 1$

Notes: Estimates are of power spectral risk measure (22) $M_\phi = \int_0^1 \gamma p^{\gamma-1} q_p dp$ where $\gamma > 1$ is the coefficient of relative risk aversion and $p$ is the cumulative probability, under the alternative assumptions that losses are distributed as: standard normal, Cauchy, standard uniform, beta(2,4) and standard Gumbel.. Results are based on numerical quadrature using trapezoidal rule with $p$ divided into 10,000 'slices'. The calculations were carried out using the CompEcon functions in MATLAB given in Miranda and Fackler (2002).

| Coefficient of Relative Risk Aversion | Standard normal | Cauchy | Standard uniform | Beta | Gumbel |
|---|---|---|---|---|---|
| 1.1 | 0.085 | 1.096 | 0.524 | 0.347 | -0.461 |
| 1.5 | 0.343 | 3.258 | 0.600 | 0.393 | -0.134 |
| 5 | 1.161 | 11.276 | 0.833 | 0.553 | 0.689 |
| 20 | 1.860 | 36.503 | 0.950 | 0.690 | 1.219 |